\documentclass[epsfig,10pt,twocolumn]{article}
\usepackage{spconf,amsmath,bm,graphicx,epsfig,citesort,setspace,times}

\renewcommand{\thetable}{\Roman{table}} % Roman numerals for tables
\newcommand{\utwi}[1]{\mbox{\boldmath $ #1$}}
\newcommand{\beq}{\begin{equation}}
\newcommand{\eeq}{\end{equation}}
\newcommand{\bea}{\begin{eqnarray}}
\newcommand{\eea}{\end{eqnarray}}
\newcommand{\bml}{\begin{mathletters}}
\newcommand{\eml}{\end{mathletters}}
\newcommand{\bc}{{\utwi{c}}}
\newcommand{\bs}{{\utwi{s}}}
\newcommand{\ba}{{\utwi{a}}}

\newcommand{\bw}{{\utwi{w}}}

\input{amssym.def}
\input{amssym}
\newcommand {\mm}[1] {\ifmmode{#1}\else{\mbox{\(#1\)}}\fi}
\newcommand{\real} {\mm{{\Bbb R}}}

\title{\bf Potential function of simplified protein models for
discriminating native proteins from decoys: Combining contact interaction and
local sequence-dependent geometry
}

\name{
Jinfeng Zhang$^1$, Rong Chen$^{1,2,3}$, Jie Liang$^{1}$
\thanks{Supported by grants from National Science
Foundation CAREER DBI0133856, DBI0078270, DMC0073601, CCR9980599
and NIH GM68958.  Phone: (312)355--1789, fax: (312)996--5921, email: {\tt jliang
@uic.edu}
}
}
\address{$^1$Department of Bioengineering, University of Illinois at Chicago, Ch
icago,IL\\
$^2$Department of Information \& Decision Science, University of Illinois at Chi
cago, Chicago,IL\\
$^3$Department of Business Statistics \& Econometrics,
Peking University,
Beijing, P.R. China
\\
{Accpeted by {\it Proc.\ 26th IEEE EMBC Conference, San Francisco, 2004}. }
}

\begin{document}
\maketitle
\setcounter{page}{1}

{\small {\em Abstract}---{\bf An effective potential function is
critical for protein structure prediction and folding simulation.
For simplified models of proteins where coordinates of only
$C_\alpha$ atoms need to be specified, an accurate potential
function is important. Such a simplified model is essential for
efficient search of conformational space.  In this work, we
present a formulation of potential function for simplified
representations of protein structures. It is based on the
combination of descriptors derived from residue-residue contact
and sequence-dependent local geometry. The optimal weight
coefficients for contact and local geometry is obtained through
optimization by maximizing margins among native and decoy
structures.  The latter are generated by chain growth and by
gapless threading.  The performance of the potential function in
blind test of discriminating native protein structures from decoys
is evaluated using several benchmark decoy sets. This potential
function have comparable or better performance than several
residue-based potential functions that require in addition
coordinates of side chain centers or coordinates of all side chain
atoms.  }}

\section{Introduction}
\label{sec:intro}

Studies of protein folding are often based on the thermodynamic
hypothesis, which postulates that the native state of a protein is
the state of lowest free energy under physiological conditions.
Based on this assumption, a potential function where the native
protein has the lowest energy is essential for protein structure
prediction, folding simulation, and protein design.

There are two steps to construct such a potential function.  The
first step is to define a proper representation of protein
structures, which is usually based on a set of numerical
descriptors characterizing the structure and sequence of the
protein.  The second step is to decide on a functional form, which
takes the descriptor vector and maps it to a real valued energy or
score for the particular structure. Frequently, potential function
$H(.)$ takes the form of $H(f(\bs, \ba)) = H(\bc) = \bw \cdot
\bc$, where the structure $\bs$ of a protein and its amino acid
sequence $\ba$ is mapped to a numerical $d$-vector $\bc$ by the
representation $f$, such that $f: (\bs, \ba) \mapsto \bc \in
\real^d$.  The potential function $ H(\bc) \in \real$ then maps
the vector $\bc$ to a real valued energy \cite{HuLiLiang03_arxiv}.
Protein representations strongly influence the effectiveness of
potential function.  The most successful potential functions rely
on detailed all-atom representation of native protein structures
with tens or hundreds thousand of parameters. Most residue-level
potential functions with significant less parameters still need
atom-level information. However, simplified representation of
proteins are essential for structure prediction and folding
simulation. Since not all heavy atoms are explicitly represented,
sampling of conformational space is far more efficient.  A
challenging open problem is whether potential function based on a
simplified protein representation can have perfect discrimination
of native or near native conformations from decoy conformations.
If so, it is important to identify the minimal representation
through which such discrimination can be achieved.

\vspace*{-0.1in}
In this work, we present a formulation of potential function designed
for simplified $n$-state representations of protein structures at
residue level.  In addition to contact interactions, we encode
sequence-dependent local geometric information.  These are combined to
form a potential function called {\it Contact and Local Geometric
Potential\/} ({\sc clgp}).  Our paper is organized as following: first
we introduce the simplified $n$-state representation of protein
structures.  We then describe the scoring function, which is a
weighted linear combination of contact and geometric parameters.  We
also discuss the optimization method to derive the parameters of the
potential function. Finally, we show the performance of {\sc clgp} on
both Park-Levitt decoy set \cite{SamudralaLevitt00_PS} and decoys
generated by gapless threading.
\vspace*{-.2in}

\section{Models and Methods}
\label{sec:meth}

\begin{figure}[tb]
\centerline{\epsfig{figure=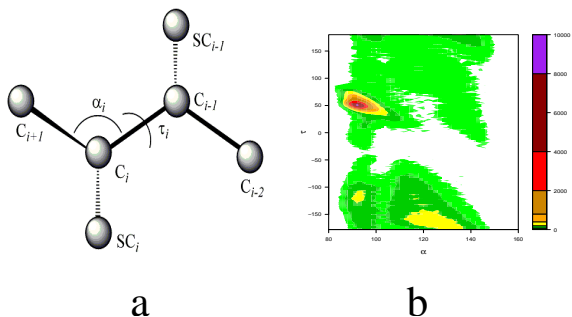,width=3.0in}}
\caption[Discrete state model]{a. Discrete state model. $\alpha_i$ and
$\tau_i$ are shown for residue $i$. Bond angle $\alpha_i$ at position
$i$ is formed by $C_{{\alpha},{i-1}}$, $C_{\alpha,{i}}$, and
$C_{\alpha,{i+1}}$. Torsion angle $\tau_i$ is the dihedral angle of
the two planes formed by atoms ($C_{i-2}$,$C_{i-1}$,$C_i$) and
($C_{i-1}$, $C_i$, $C_{i+1}$);
b. Distribution of $\alpha$ and $\tau$ angles.}
\label{Fig:model}
\end{figure}

{\sc Representation of protein structures.} Our representation of
protein structures is an off-lattice discrete state model
\cite{ParkLevitt96_JMB}. Specifically, we represent all backbone atoms
by the $C_\alpha$ atom, and all side chain atoms by one additional
atom attached to the main chain $C_\alpha$ atoms
(Figure~\ref{Fig:model}a).  There are totally 20 different atom types
(1 $C_\alpha$ and 19 side chain atoms).  The backbone structure is
described by the bond angle $\alpha$ and torsion angle $\tau$ at each
$C_\alpha$ position (Figure~\ref{Fig:model}a).  The overall three
dimensional structure of a protein of length $N$ is completely
determined by the set of the bond angles and torsion angles
$\{(\alpha_i, \tau_i)\}, \quad 1<i<N $, where $i$ represents the
$i$-th position of the backbone.  The distance between each $C_\alpha$
atom and its side chain atom is predefined for each residue type.

In discrete state model, except residues at the two ends each main
chain position can take $n$ discrete pairs of ($\alpha,\tau$) angles,
which is the states at this position. To determine the optimal number
of states and the exact values of the ($\alpha,\tau$) angles for the
discrete representation, we examine the distribution of $\alpha$ and
$\tau$ angles calculated from 980 non-homologous proteins from {\sc
PDBSelect}. Analogous to the Ramachandran plot, the distribution of
($\alpha,\tau$) also has densely and sparsely populated regions, which
corresponds to different secondary structure types
(Fig~\ref{Fig:model}b). And the distribution differs for different
amino acids. To obtain exact values for each discrete states, we use
$k$-mean clustering to group $\alpha$ and $\tau$ values observed in
native proteins into $k$ from 3 to 10 clusters for each amino acid
residue type. The cluster centers are then taken as the discrete
states.

\begin{figure}[tb]
\centerline{\epsfig{figure=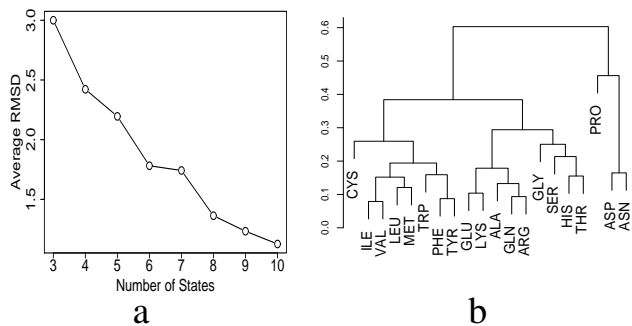,width=3.5in}}
\caption[Clustering of amino acids]{a. Average RMSD of the best discrete state models fitted to PDB structures of 348 proteins from 3 to 10 state;
b. Hierarchical clustering of amino acids by local sequence geometry propensity. }
\label{Fig:aa_clus}
\end{figure}

The discrete state representation is associated with discrete
conformational space that is different from the continuous space of
PDB structures. To represent PDB structure in the simplified discrete
space, we need to map an PDB structure to a structure in the discrete
space, with the requirement that these two structures are most similar
by some criteria. In this study we consider two such criteria: global
structural similarity and local structural similarity. To generate a
discrete structures $\bs_d$ that are globally similar to a PDB
structure $\bs$, we use a heuristic ``build-up'' algorithm
\cite{ParkLevitt95_JMB}, which globally fits the discrete state model
$\bs_{\min}$ to the PDB structure $\bs$, namely, $\bs_{\min} =
\arg_{\bs_d} \min \left[ \rm{RMSD}(\bs, \bs_d)\right]$.  We obtain
discrete best-fit structure for 348 proteins, with length ranging from
40 to more than 1,000. The average RMSD to the PDB structure of the
best-fit structures is 2.4 \AA\ for 4-state model, 1.8 \AA\ for
6-state, and 1.1 \AA for 10-state. Figure~\ref{Fig:aa_clus}a shows the
average RMSD for models with 3 to 10 states. The relative high quality
of these fitted discrete state structures indicates that a model with
four to six states is sufficient to generate near native structures
(NNS) that have small RMSD to experimental structures. In this study,
we use 4-state model for all amino acid residues.  To generate
discrete state model with local similarity to PDB structures, each
residue is simply assigned a discrete state most similar to its local
($\alpha$,$\tau$) angle in the PDB structure.

{\sc Clustering amino acids by first order state transition
propensity.}  We reduce the alphabet of twenty amino acids to a
simpler alphabet based on their local geometric
properties. Simplifying the residue alphabet by geometric properties
helps to improve characterization of the local relationship between
sequence and backbone structure, and alleviate the problem of
inadequate data for deriving empirical potential functions.

A protein structure can be represented uniquely by a sequence of
$(a,x)$ pairs, where $a$ is amino acid residue type and $x$ is the discrete
state. For four discrete state and 20 amino acid types, the total
number of possible descriptors at one residue position is
$4\times20=80$. For simplicity, we use $s \in [1...80]$ to represent
the state (discrete state and amino acid type) a residue may take.  We
define first order state transition probability as: $p_{s_1,s_2} =
p[s_2|s_1] = p[(a_{2},x_{2})|(a_1,x_1)].  $ First order state
transition geometric score can then be defined as: $
s_g=\ln{\frac{p_{s_1,s_2, obs}}{p_{s_1,s_2,exp}}}, $ where
$p_{s_1,s_2,obs}$ is the observed probability of transition from
state $s_1$ to state $s_2$, and $p_{s_1,s_2,exp}$ is the expected
transition probability from state $s_1$ to $s_2$ for a random
distribution, which is the product of the relative frequency of $a_1$,
$x_1$, $a_2$, and $x_2$.  We cluster twenty amino acids using their
first order state transition score.  Specifically, we define the
distance between two amino acids as the distance between the
corresponding row vectors in the state transition matrix
(Figure~\ref{Fig:aa_clus}b).

This clustering approach based on local sequence-structure
relationship provides a useful method to group amino acids and to
obtain a simplified protein alphabet. For this study, we take a
simplified alphabet of 5 letters: \{C,I,V,L,M,W,F,Y\},
\{A,E,K,Q,R\}, \{G,S,H,T\}, \{P\}, and \{D,N\} as our simplified
alphabet.  Each residue position now has $4 \times 5 = 20$
possible descriptors: there are 4 discrete $(\alpha, \tau)$ states
and 5 types of amino acids.

The representation of a protein structure is now a vector that
includes both contact interaction (with 210 different types of
contacts) and local geometric score (with $20 \times 20 = 400$
parameters for each of the $4\times5$ geometric descriptors for two
consecutive residues).

{\bf \sc Combining contact and local geometric descriptors.  } The
form of the scoring function in this study is a weighted linear
combination of all descriptors: $ H(c) = \bw \cdot \bc$, where $\bc$
is the descriptor vector, $\bw$ is the weight vector.  We obtain
weight vector $\bw$ using optimization method. For a linear scoring
functions, the basic requirement based on Anfinsen experiment is:
\[
 \bw \cdot (\bc_N - \bc_D) + b < 0, \quad \rm{where}\ b
\geq 0,
\]
where $\bc_N$ and $\bc_D$ are descriptor vectors for native and decoys
structures of a specific protein, respectively, and $b$ is the energy
gap required between a native and decoy structure. Each pair of native
descriptor vector and decoy descriptor vector provides one inequality
constraint. Together many such constraints define a convex polyhedron
$\cal P$ where feasible solutions $\bw$ vectors are located. For nonempty $\cal
P$,  there could be infinite number of choices of
$\bw$, all with perfect discrimination.
The optimal weight vector $\bw$ can be found by solving the following
quadratic programming problem:
\begin{eqnarray}
\mbox{Minimize } & \frac{1}{2} || \bw||^2
\\
\mbox{subject to} & \bw \cdot (\bc_N - \bc_D) + b < 0 \mbox{ for all }
N  \mbox{ and } D.
\label{Eqn:PrimalLinear}
\end{eqnarray}
The solution maximizes the distance $b/||\bw||$ of the plane $(\bw,
b)$ to the origin. Here, we use
a support vector machine to find such
an optimal  $\bw$.

We take 980 non-homologous proteins from the {\sc PDBSelect} database.
Among these, 652 proteins are randomly chosen as training set, and the
remaining 328 proteins are used as testing set.  We first use gapless
threading to generate $>$10 millions of decoys to train and test our
contact and local geometric potential function ({\sc clgp}).  This is
then complemented by training using explicit decoys obtained from \cite{LooseFloudas04_P}.

\vspace*{-0.3in}
\section{Results}
{\sc Performance on gapless threading decoys.}  The performance of the
potential function on gapless threading decoys is listed in
Table~\ref{tab:glt_pfm}. Among 328 test proteins, only 4 proteins are
misclassified. The accuracy of {\sc clgp} is near 99\%.  We compare
{\sc clgp} with several other residue based potential functions,
including those developed by Tobi \& Elber {\it et.al.} (TE13)
\cite{TobiElber00_P}, Miyazawa \& Jernigan (MJ)
\cite{MiyazawaJernigan96_JMB}, and Bastolla \& Vendruscolo {\it
et. al.} (BV) \cite{BastollaVendruscolo01_P}. Although these potential
functions are residue based potential function, they still require
all-atom representation since they either need to calculate the side
chain centers or need to compute explicit atom-atom contacts. {\sc
clgp} is the only potential function applicable to representations
with only $C_\alpha$ atoms.  It has better performance than other
residue based potential and is comparable to that of all-atom
potential.

\begin{table}[tb] % one column table
\begin{small}
\begin{center}
\caption[performance on GLT]{\label{tab:glt_pfm} 
The number of mis-classifications
using {\sc clgp} and other residue based potential functions.
$C_\alpha$: Computation of potential function need only $C_\alpha$
atoms. SC: Computation of potential function needs side chain
center. AT: Computation of potential function needs all-atom
representation.} 
\vspace*{.1in}
\begin{tabular}{|r|r|r|}
\hline
Potential   & Complexity      & Mis-classified \\
Functions    &                 & Proteins       \\
\hline 
{\sc clgp}        &  $C_\alpha$     & 4/328          \\
\hline TE13        &  SC            & 7/194          \\
\hline BV          &  AT             & 2/194          \\
\hline MJ          &  SC            & 85/194         \\
\hline
\end{tabular}
\vspace*{-0.2in}
\end{center}
\end{small}
\end{table}

{\sc Performance on other decoy sets.}  Potential function trained
using gapless threading decoys usually do not works well for more
sophisticated decoys generated by other methods. To further train the
{\sc clgp} potential function, we include additional decoy set generated by
Loose {\it et. al.}  \cite{LooseFloudas04_P}, which contains $>80,000$
decoy structures generated by energy minimization protocols for a set
of 829 proteins.  The new {\sc clgp} potential function is tested using
the Park-Levitt decoy set create in \cite{SamudralaLevitt00_PS}. The
performance of our potential function on several decoy sets is shown
on table~\ref{tab:dec_pfm}. We also compared our results with three other
residue-based potential function, namely, TE13, LHL (Li, Hu, \& Liang
\cite{LiHuLiang03_P}) , and MJ. Performance of {\sc clgp} in general
is better or comparable to other potential functions. It performs
better than other potential functions on {\sc lmds}
decoy set.

\begin{table}[tb] % one column table
\begin{small}
\begin{center}
\caption{\label{tab:dec_pfm} 
Performance of {\sc clgp} for ParkLevitt decoy set.}
\vspace*{.1in}
\begin{tabular}{|l|r|r|r|r|r|}
\hline 
Decoy sets  & {\sc clgp}           & TE13      & LHL & MJ \\ \hline 
Complexity  & $C_\alpha$    & SC  & AT   & SC    \\ \hline
A) 4state \cite{ParkLevitt96_JMB}   &           &       &    & \\
1ctf    & 1 & 1 & 1 & 1 \\
1r69    & 1 & 1 & 1 & 1 \\
1sn3        & 2             &  6        & 1     &  2  \\
2cro        & 2             &  1        & 1     &  1  \\
3icb        & 1             &  --       & 5     & --  \\
4pti        & 2             &  7        & 1     & 3   \\
4rxn        & 3             & 16        & 51    & 1   \\ \hline
B) lmds \cite{Levitt83_JMB}    &         &    &    &  \\
1b0n-B      & 1             & --        & 2     & --  \\
1bba        & 436           & --        & 217   & --  \\
1ctf & 1 & 1 & 1 & 1 \\
1fc2        & 83            & 14        & 500   & 501 \\
1dtk        & 1             & 5         & 2     & 13  \\
1igd        & 1             & 2         & 9     & 1   \\
1shf-A      & 3             & 1         & 17    & 11  \\
2cro & 1 & 1 & 1 & 1 \\
2ovo        & 4             & 1         & 3     & 2   \\
4pti        & 1             & --        & 9     & --  \\ \hline
 C) lattice\_ssfit \cite{SamudralaLevitt99_PSB,XiaLevitt00_JCP} &   &  & &   \\
1beo        & 1             & --        &  1    & --  \\
1ctf & 1 & 1 & 1 & 1 \\
1dkt-A      & 1             & 2         &  1    & 32  \\
1fca        & 7             & 36        &  40   & 5   \\
1nkl & 1 & 1 & 1 & 1 \\
1trl-A      & 56            & 1         &  5    & 4   \\
1pgb & 1 & 1 & 1 & 1 \\
4icb        & 1             & --        & 1     & --  \\ 
\hline
\end{tabular}
\end{center}
\end{small}
\end{table}

\section{Summary and Conclusion}

In this study, we have developed an effective potential function for
simplified representation of protein structures. By $k$-mean
clustering, we obtained accurate discrete states and demonstrated the
accuracy of model generated by these states using a modified build-up
algorithm. We then simplified amino acid alphabet using their local
sequence-structure relationship. The first order state transition
propensity is constructed for this purpose.  This potential function
combines both contact interaction and local sequence-structure
relationship. Contact interaction correlates well with more global
topological information, while local sequence-structure relationship
contains local geometric information. We find native structures can be
well stabilized against decoy structures. The performance of this
potential function is better than or comparable to other potential
functions but it requires significantly less detailed description.  We
expect this potential function to be useful in protein structure
prediction and fold simulation of simplified protein models.

\bibliography{prelim,lattice,pack,design,potential,pair}
\bibliographystyle{unsrt}

\end{document}